\documentclass{Interspeech}

\interspeechcameraready 

\hypersetup{
    colorlinks=true,
    citecolor=[HTML]{006EBB},
    linkcolor=[HTML]{006EBB},
    urlcolor=[HTML]{006EBB}
}

\title{EzAudio: Enhancing Text-to-Audio Generation with Efficient Diffusion Transformer}
\author[affiliation={1}]{Jiarui}{Hai$^{*}$}
\author[affiliation={2}]{Yong}{Xu}
\author[affiliation={2}]{Hao}{Zhang}
\author[affiliation={2}]{Chenxing}{Li}
\author[affiliation={1}]{Helin}{Wang}
\author[affiliation={1}]{Mounya}{Elhilali}
\author[affiliation={2}]{Dong}{Yu}

\affiliation{Department of Electrical and Computer Engineering}{Johns Hopkins University}{MD, USA}
\affiliation{AI Lab}{Tencent Americas}{WA, USA}
\email{jhai2@jhu.edu, lucayongxu@global.tencent.com}


\usepackage{comment}

\usepackage{url}
\usepackage{multirow}
\usepackage{booktabs, tabularray}
\usepackage[table,xcdraw]{xcolor}
\usepackage{tablefootnote}
\definecolor{lightgreen}{HTML}{DFF4DB} 
\definecolor{lightblue}{HTML}{D7E6FB} 
\definecolor{lightpurple}{HTML}{EDD9F0} 
\definecolor{lighterpurple}{HTML}{F6ECF8}
\definecolor{lightestpurple}{HTML}{FBF6FC}
\usepackage{tabularx} 

\usepackage{threeparttable}
\usepackage{threeparttablex} 

\begin{document}

\maketitle

\renewcommand{\thefootnote}{\fnsymbol{footnote}}
\footnotetext[1]{Work done during J. Hai's internship at Tencent AI Lab, USA.}
\renewcommand{\thefootnote}{\arabic{footnote}}

\begin{abstract}
We introduce EzAudio, a text-to-audio (T2A) generation framework designed to produce high-quality, natural-sounding sound effects. Core designs include: (1) We propose EzAudio-DiT, an optimized Diffusion Transformer (DiT) designed for audio latent representations, improving convergence speed, as well as parameter and memory efficiency. (2) We apply a classifier-free guidance (CFG) rescaling technique to mitigate fidelity loss at higher CFG scores and enhancing prompt adherence without compromising audio quality.  (3) We propose a synthetic caption generation strategy leveraging recent advances in audio understanding and LLMs to enhance T2A pretraining. We show that EzAudio, with its computationally efficient architecture and fast convergence, is a competitive open-source model that excels in both objective and subjective evaluations by delivering highly realistic listening experiences.
\end{abstract}

\section{Introduction}
The rapid advancement of diffusion-based generative models has transformed content creation, particularly in image synthesis \cite{rombach2022high}. Inspired by this, early text-to-audio (T2A) methods used spectrogram-based representations, evolving into a powerful approach for high-quality sound generation \cite{ghosal2023tango, liu2023audioldm, huang2023make}. Recent work \cite{evans2024long, huang2023make2, vyas2023audiobox, haji2024taming} has improved T2A quality by adopting one-dimensional (1D) latent audio representations. The Diffusion Transformer (DiT) \cite{peebles2023scalable}, leveraging Adaptive LayerNorm (AdaLN) for diffusion step fusion, has shown strong performance in visual generation and, more recently, in sound generation \cite{evans2024long}. Despite these advances, recent T2A pipelines still have room for improvement: (1) DiT in audio generation requires substantial memory and training costs and could benefit from further optimization for T2A tasks and latent audio representations. (2) We find the T2A model using waveform latents exhibit noticeable fidelity loss at high classifier-free guidance (CFG) \cite{ho2021classifier} scores. While higher guidance improves prompt coherence, it can distort the waveform amplitude distribution, subsequently affecting frequency features and introducing artifacts, leading to degradation in generation quality.

Beyond model design, pretraining plays a crucial role in achieving high-quality T2A due to the scarcity of human-labeled data. Strategies \cite{vyas2023audiobox, chenpixart} have been proposed to leverage unlabeled data for representation learning and improving generation quality. However, text-to-audio mapping pretraining strategies still face challenges. Using CLAP embeddings \cite{liu2023audioldm} derived from unlabeled audio data and switching to text-derived CLAP embeddings in downstream tasks can limit performance due to mismatches between audio and text representations. Tagging-based pseudo captions \cite{ghosal2023tango, mei2024wavcaps, vyas2023audiobox} directly incorporate text conditions during pretraining but lack sequential information about sound events, limiting the model’s ability to process fine-grained prompts in downstream tasks. Synthetic audio data \cite{huang2023make, huang2023make2} offers precise descriptions and timing alignment but is difficult to prepare and may introduce artifacts or unnatural characteristics due to discrepancies with real audio.


To address these challenges, we propose following core designs: (1) \textbf{EzAudio-DiT}, an optimized DiT architecture for efficient, high-quality T2A. It features a novel AdaLN variant that reduces parameters and memory consumption without compromising performance, along with long-skip connections to accelerate convergence. (2) We enhance CFG sampling by adopting \textbf{CFG rescaling} \cite{lin2024common}, originally developed to prevent over-exposure in image generation. We demonstrate that when applied to waveform latents, it mitigates fidelity degradation at high CFG scores while preserving strong prompt adherence, eliminating the need for meticulous CFG tuning. (3) We leverage recent advances in audio understanding and LLMs to generate high-quality \textbf{synthetic caption} data for efficient T2A pretraining. Specifically, we prepare the following sources of synthetic caption data: (a) generating captions using audio captioning and audio-language models, which have demonstrated the ability to interpret complex auditory scenes \cite{chu2023qwen, ghosh2024gama, kong2024improving}, and (b) enriching strong sound event labels \cite{hershey2021benefit} with LLMs to generate captions that provide detailed sequential information about sound events.

As a result of these designs, \textbf{EzAudio}\footnote{Code and demo: \href{https://haidog-yaqub.github.io/EzAudio-Page/}{https://haidog-yaqub.github.io/EzAudio-Page/}} achieves fast convergence with reduced parameters and memory usage and is able to generate highly realistic audio. It stands out as a competitive open-source model in both objective and subjective evaluations. We hope our model and pipeline empower researchers and startups to develop T2A models more effectively and efficiently.

\section{Method}

\begin{figure*}[t]
  \centering
  \includegraphics[width=0.95\textwidth]{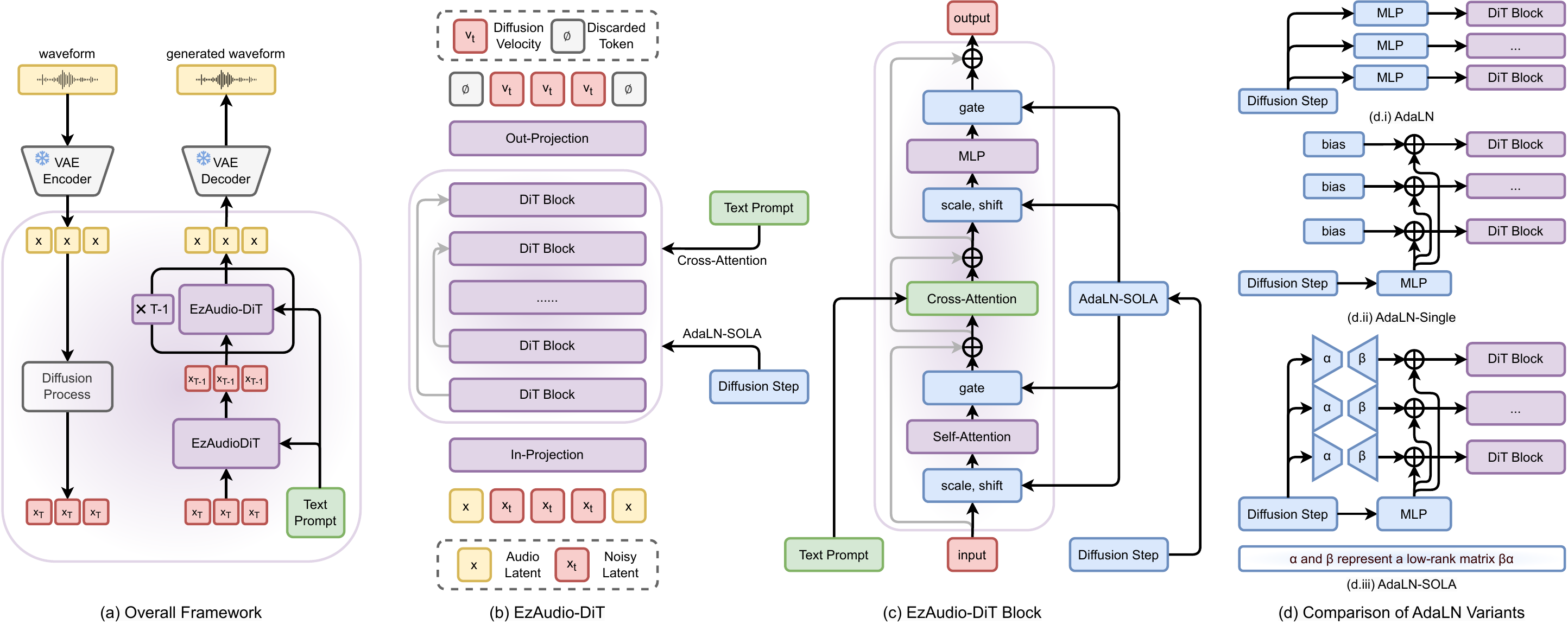}
  \caption{The framework of EzAudio and the architectural details of EzAudio-DiT.}
  \label{fig:method}
\end{figure*}


EzAudio builds on recent advances in diffusion-based audio and music generation \cite{ghosal2023tango, evans2024long}. As shown in Figure \ref{fig:method}, it comprises three components: (1) a FLAN-T5-based text encoder \cite{chung2024scaling} for processing audio descriptions, (2) a latent diffusion model for generating audio latents, and (3) a waveform VAE \cite{evans2024long} for reconstructing waveforms from audio latents.

The following sections detail EzAudio's core designs: Section \ref{model} introduces the proposed EzAudio-DiT for diffusion modeling, Section \ref{rescale} describes CFG rescaling for sampling, and Section \ref{train} covers data labeling and training strategy.

\subsection{Efficient EzAudio-DiT}
\label{model}
Stable Audio \cite{evans2024long} has successfully used DiT \cite{peebles2023scalable} for text-to-music generation. However, we find opportunities to optimize DiT's efficiency and convergence speed in audio generation. To achieve this, we propose two key modifications that enhance parameter and memory efficiency while accelerating convergence, without compromising training stability. These designs include:

\textbf{AdaLN-SOLA:} The AdaLN layers in DiT are crucial for managing both image class conditions and diffusion steps but account for a significant portion of the model’s parameters. However, in T2A, where cross-attention processes text conditions, AdaLN can be simplified. AdaLN-Single \cite{chenpixart} addresses this by sharing a single AdaLN module across all DiT blocks but degrades performance and even destabilizes training. To address this, we propose \textbf{AdaLN-SOLA} (\textbf{AdaLN}-\textbf{S}ingle \textbf{O}rchestrated by \textbf{L}ow-rank \textbf{A}djustment), inspired by low-rank adaption methods\cite{hu2021lora}. As shown in Figure~\ref{fig:method}~(d), AdaLN-SOLA retains a shared AdaLN module but incorporates block-specific low-rank matrices that dynamically adjust it based on diffusion steps. This approach reduces parameters and memory usage while preserving performance and stability.

\textbf{Long-skip Connection:} Earlier diffusion models use long-skip connections to propagate low-level features and diffusion steps into deeper layers for effective modeling.  Recent DiT architectures \cite{chenpixart, evans2024long} remove these connections, relying on transformer residuals for feature propagation and assuming AdaLN can handle the diffusion steps. However, we find that removing skip connections slows convergence and degrades performance, particularly for waveform latent embeddings with 128 channels—far more than typical image representations—making them difficult to process solely through residual connections. To address this, we integrate long-skip connections into DiT, inspired by U-ViT designs \cite{bao2023all, le2024voicebox, vyas2023audiobox}, allowing low-level features to directly reach deeper transformer blocks, as shown in Figure~\ref{fig:method}~(b).

\subsection{CFG Rescaling}\label{rescale}
The CFG \cite{ho2021classifier} is used to direct the diffusion sampling. It modifies the output \(v\) only during the reverse process according to:
\begin{flalign}
\textstyle v_{cfg} = v_{neg} + w (v_{pos} - v_{neg}),
\end{flalign}
where \(w\) is the guidance scale, and \(v_{pos}\) and \(v_{neg}\) represent model outputs under positive and negative prompts, with \(v_{cfg}\) being the adjusted velocity. By default, the negative prompt is set to empty, corresponding to the unconditional case. 

A higher guidance scale enhances prompt alignment but can disrupt the waveform's amplitude distribution, affecting frequency characteristics and ultimately degrading generation quality. The CFG rescaling technique \cite{lin2024common} is used to adjust the magnitude of \(v_{cfg}\) while preserving its direction when a large \(w\) is employed.
\begin{flalign}
\textstyle v_{re} = v_{cfg} \cdot \mathrm{std}(v_{pos}) \cdot \mathrm{std}(v_{cfg})^{-1}, \\
\textstyle v'_{cfg} = \phi \cdot v_{re} + (1-\phi) \cdot v_{cfg},
\end{flalign}
where \(\phi\) is the rescaling factor, with \(v'_{cfg}\) denoting the refined CFG velocity for diffusion sampling.

\subsection{Training Strategy}\label{train}
\textbf{Synthetic Caption Data Generation}: We utilize multiple approaches to generate synthetic caption data, enhancing caption diversity and richness: \textbf{(1) Auto-ACD} \cite{sun2024auto} utilizes audio and video captioning models to generate initial captions, which are then refined by a language model into natural audio descriptions for AudioSet and VGGSound. \textbf{(2) AS-Qwen-Caps} uses Qwen-Audio\footnote{We compare Qwen-Audio \cite{chu2023qwen} and GAMA \cite{ghosh2024gama}, selecting Qwen-Audio for its higher accuracy and fewer hallucinations on AudioCaps.} \cite{chu2023qwen}, one of the leading audio-language models, to describe audio from AudioSet and VGGSound; \textbf{(3) AS-SL-GPT4-Caps} uses OpenAI's GPT-4o-mini API\footnote{\href{https://platform.openai.com/docs/models/gpt-4}{https://platform.openai.com/docs/models/gpt-4}} to prepare descriptions that emphasize sequential information based on temporal annotations from the strongly labeled subset of AudioSet \cite{hershey2021benefit}. To ensure the quality and accuracy of captions, we use a CapFilt-like \cite{li2022blip, kong2024improving} filtering method, leveraging a pre-trained CLAP model \cite{wu2023large} to discard audio-caption pairs with similarity scores below a set threshold.

\textbf{Multi-Stage Training}: We adopt a three-stage training approach \cite{chenpixart, vyas2023audiobox} to leverage unlabeled audio data and enhance generation quality. \textbf{(1) Masked Audio Modeling:} Following diffusion-based mask pretraining methods \cite{le2024voicebox, vyas2023audiobox, gao2023masked}, the diffusion model is first trained to predict masked tokens from unmasked ones, without text conditioning. A random portion of tokens—ranging from 25\% to 100\% with a minimum span of 0.2s—is masked, and the cross-attention module in transformer blocks is excluded during this stage. \textbf{(2) Text-Audio Alignment:} This stage integrates synthetic captions to facilitate text-audio alignment learning. Building on the masked modeling stage, we introduce a randomly initialized cross-attention module into each DiT block to process text conditions. To ensure a smooth training transition, we initialize the output projection layer of the cross-attention module to zero. Additionally, to encourage greater reliance on text input, we set a fixed 75\% probability of fully masking all tokens during training. \textbf{(3) Supervised Fine-Tuning:} Finally, following Tango \cite{ghosal2023tango}, we fine-tune the model on AudioCaps \cite{kim2019audiocaps} to further enhance performance.

\section{Experiments}
\subsection{Experimental Setups}\label{exp}
We conduct experiments using a 24kHz sample rate for both the waveform VAE and the T2A model. The waveform latent operate at 50Hz and consists of 128 channels. We train the waveform VAE on AudioSet \cite{gemmeke2017audio} for 1 million steps, enabling it to handle a wide variety types of sounds. For DiT models, DiT-L consists of 24 DiT blocks, each with 1024 channels, while DiT-XL has 28 DiT blocks, each with 1152 channels. The rank in AdaLN-SOLA is 32 for DiT-L and 36 for DiT-XL.
The LDM employs velocity ($v$) prediction and Zero-SNR schedulers \cite{lin2024common}, both effective in diffusion-based image and audio generation \cite{podell2023sdxl, hai2024dpm}. 
We use 50 sampling steps and a CFG score of 3 by default in the ablation studies presented in Sections \ref{exp.dit} and \ref{exp.train}.

Following previous T2A studies \cite{liu2023audioldm, liu2024audioldm, ghosal2023tango, huang2023make2}, we evaluate our model using Frechet Distance (FD)\footnote{We exclude FAD due to reliability concerns \cite{liu2023audioldm, kong2024improving}.}, Kullback–Leibler (KL) divergence, and Inception Score (IS), with pre-trained PANNs \cite{kong2020panns} as the feature extractor. Additionally, we employ CLAP\footnote{\href{https://huggingface.co/laion/larger_clap_general}{https://huggingface.co/laion/larger\_clap\_general}} \cite{wu2023large} to assess the coherence between the generated audio and the text prompt. All audio samples are \textbf{resampled to 16kHz} during evaluation. The AudioCaps test set, comprising 900 audio clips with 882 currently available, is used for evaluation. Each clip has five captions, and we \textbf{randomly select one caption per clip}, following AudioLDM and Tango \cite{ghosal2023tango, liu2023audioldm}.

\begin{figure}[t]
  \centering
  \includegraphics[width=0.45\textwidth]{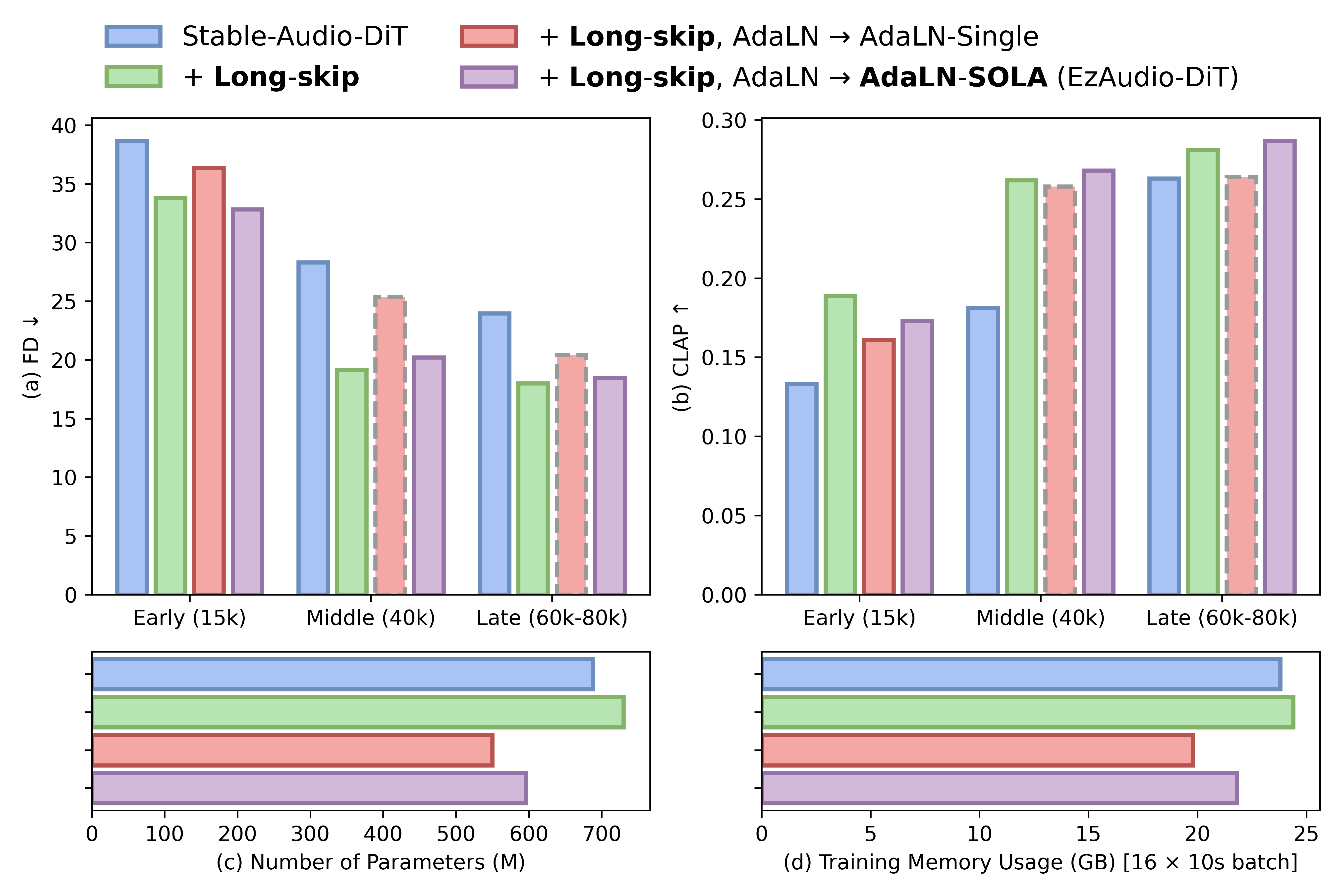}
\caption{Ablation of DiT Design. The \textcolor{gray}{gray dashed} edges indicate results from training resumed after a crash.}
  \label{fig:design}
\end{figure}


\subsection{Ablation of DiT Architecture}\label{exp.dit}

We perform an ablation study on different DiT designs using the AudioCaps dataset, training for 80k steps with a batch size of 128 and a learning rate of 1e-4, following the DiT-L configuration in Section~\ref{exp}. Stable-Audio-DiT~\cite{evans2024long}, which extends DiT~\cite{peebles2023scalable} with cross-attention and Rotary Position Embedding (RoPE)~\cite{su2024roformer}, serves as our baseline. We investigate the effects of adding long-skip connections and replacing AdaLN with either AdaLN-Single or the proposed AdaLN-SOLA. We compare convergence across three training stages. Model performance improves steadily during the early and middle stages, with results reported at 15k and 40k steps. In the late stage, performance stabilizes with minor fluctuations, and the best scores between 60k and 80k steps are reported based on validation loss.

The key findings can be summarized as follows: (1) As shown in Figure~\ref{fig:design} (a) and (b), long-skip connections significantly accelerate convergence and lead to better model performance; (2) Replacing AdaLN with AdaLN-Single leads to performance degradation and introduces numerical instability, causing training crashes, whereas AdaLN-SOLA maintains stability with minimal impact on performance; (3) Figure~\ref{fig:design} (c) and (d) illustrates that long-skip connections slightly increase model parameters and memory usage, while AdaLN-SOLA substantially reduces both, resulting in a more lightweight model.  In conclusion, EzAudio-DiT achieves faster convergence and greater efficiency than Stable-Audio-DiT.



\begin{table}[t]
    \caption{Comparison of pretraining methods.}
    \centering
    \scriptsize
    \resizebox{0.45 \textwidth}{!}{
    \begin{tabular}{c c | c c c c}
            \toprule
            \textbf{Dataset} & \textbf{Mask Mod.} & \textbf{FD$\downarrow$} & \textbf{KL$\downarrow$} & \textbf{IS$\uparrow$} & \textbf{CLAP$\uparrow$} \\
            \midrule
            WavCaps & No & 17.79 & 1.66 & 9.60 & 0.273 \\
            \textbf{EzAudioCaps} & No & 16.60 & 1.67 & 10.04 & 0.288 \\
            \midrule
            \textbf{EzAudioCaps} & \textbf{Yes} & \textbf{15.46} & \textbf{1.44} & \textbf{10.11} & \textbf{0.294} \\
            \bottomrule
        \end{tabular}
    }
    \label{tab:method}
\end{table}

\begin{table}[t]
    \caption{Ablation of CLAP filtering.}
    \centering
    \scriptsize
    \resizebox{0.45 \textwidth}{!}{
\begin{tabular}{c c | c c c c}
    \toprule
     \textbf{Threshold} & \textbf{\# Samples} & \textbf{FD$\downarrow$} & \textbf{KL$\downarrow$} & \textbf{IS$\uparrow$} & \textbf{CLAP$\uparrow$} \\
    \midrule
     0.35 & 0.58M & \underline{16.17} & 1.48 & 9.85 & 0.290 \\
     0.45 & 0.11M & 16.27 & \textbf{1.40} & \textbf{10.31} & \textbf{0.303} \\
    \midrule
     0.40 & 0.27M & \textbf{15.46} & \underline{1.44} & \underline{10.11} & \underline{0.294} \\
    \bottomrule
\end{tabular}

    }
    \label{tab:filter}
\end{table}
 
\begin{table*}[t]
    \centering
    \caption{Comparison of EzAudio and T2A models with evaluation results on AudioCaps. \textdagger\ denotes trainable parameters.}
    \scriptsize
    \setlength{\tabcolsep}{6pt} 
    \begin{tabular}{l | c c c c | c c c c}
        \toprule
        \textbf{Method} & \textbf{Model} & \textbf{\# Params.\textdagger} & \textbf{Pretrain Data} & \textbf{Text Encoder} & \textbf{FD$\downarrow$} & \textbf{KL$\downarrow$} & \textbf{IS$\uparrow$} & \textbf{CLAP$\uparrow$} \\
        \midrule
        Ground Truth & -- & -- & -- & -- & -- & -- & -- & 0.302 \\
        \midrule
        Tango \cite{ghosal2023tango} & 2D U-Net & 866M & Synthetic Caption & FLAN-T5 & 19.07 & 1.33 & 7.70 & 0.293 \\
        Tango-AF \cite{kong2024improving} & 2D U-Net & 866M & Synthetic Caption & FLAN-T5 & 21.84 & \underline{1.32} & 9.20 & 0.269 \\
        AudioLDM\footref{fn:audioldm} \cite{liu2023audioldm} & 2D U-Net & 739M & CLAP Embedding & CLAP & 30.96 & 2.36 & 7.38 & 0.197 \\
        AudioLDM-2\footref{fn:audioldm} \cite{liu2024audioldm} & 2D U-Net & 712M & Synthetic Caption & CLAP + FLAN-T5 & 25.03 & 1.75 & 8.13 & 0.236 \\
        Make-An-Audio \cite{huang2023make} & 2D U-Net & 453M & Synthetic Audio & CLAP & 18.77 & 1.71 & 8.80 & 0.244 \\
        Make-An-Audio-2\footref{fn:maa} \cite{huang2023make2} & 1D Transformer & 937M & Synthetic Audio & CLAP + FLAN-T5 & 16.16 & 1.42 & 9.93 & 0.284 \\
        Gen-AU-Large  \cite{haji2024taming} & 1D Transformer & 1.25B & Synthetic Caption & CLAP + FLAN-T5 & 17.21 & 1.40 & \textbf{11.42} & \textcolor{gray}{0.270}\rlap{\footref{fn:gen-au}}  \\
        \midrule
        EzAudio-L & 1D Transformer & 596M & Synthetic Caption & FLAN-T5 & \underline{15.59} & 1.38 & 11.35 & \textbf{0.319} \\
        EzAudio-XL & 1D Transformer & 874M & Synthetic Caption & FLAN-T5 & \textbf{14.98} & \textbf{1.29} & \underline{11.38} & \underline{0.314} \\
        \bottomrule
    \end{tabular}
        \label{tab:merged_comparison}
\end{table*}
    
 \subsection{Ablation of Training Strategy}\label{exp.train}

We conduct an ablation study comparing our dataset with WavCaps \cite{mei2024wavcaps}, which enriches audio tags using ChatGPT but often lacks sequential information\footnote{Comparison with our dataset available on the Demo page.} and has been used for pretraining in Tango-Full \cite{ghosal2023tango}. Additionally, we evaluate the impact of mask modeling. For pretraining without mask modeling, we use a batch size of 128 and train for 150K steps at a 1e-4 learning rate with synthetic caption data, followed by 30K fine-tuning steps on AudioCaps at 1e-5. When incorporating mask modeling, we first train on AudioSet for 100K steps at 1e-4, then perform 50K steps on synthetic caption data at 5e-5, and conclude with 30K fine-tuning steps at 1e-5.

As shown in Table \ref{tab:method}, our proposed dataset improves generation quality and strengthens text coherence. Also, mask modeling pretraining further enhances  overall generation quality.

Additionally, we evaluate different thresholds for filtering synthetic captions. The threshold selection is based on the mean CLAP score of AudioCaps, which is around 0.30. We set higher thresholds than this number to prioritize quality. All models are trained with mask pretraining. As shown in Table \ref{tab:filter}, a lower threshold allows for more diverse but noisier data, negatively impacting all metrics, whereas a higher threshold improves most metrics but reduces FD and limits data diversity. We adopt a threshold of 0.40 for EzAudioCaps, as it provides the best balance between data diversity and model performance.

\subsection{Ablation of CFG and CFG Rescaling}\label{exp.noise}

\begin{figure}[t]
  \centering
  \includegraphics[width=7.5cm]{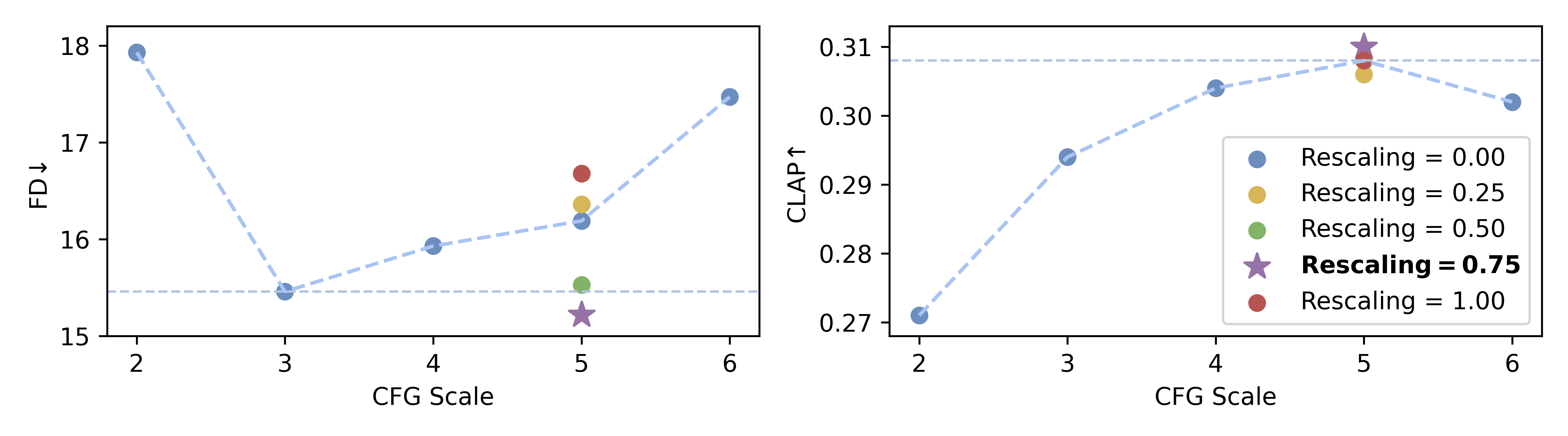}
  \caption{Ablation of CFG scales and rescaling factors.}
  \label{fig:cfg}
\end{figure}

As shown in Figure \ref{fig:cfg}, we evaluate CFG scores using the model trained in Section \ref{exp.train}. Higher CFG values improve text-audio alignment but also increase FD, indicating a decline in audio quality. With CFG = 5 yielding the highest CLAP score and less FD degradation, we apply rescaling at this level.  A rescaling factor around 0.50–0.75 helps maintain strong prompt alignment while mitigating the negative impact on audio quality.

\subsection{Comparison with State-of-the-art}

We compare EzAudio with recent open-source T2A models, introducing two variants: EzAudio-L and EzAudio-XL, which differ only in model size, corresponding to EzAudio-DiT in L and XL configurations, as described in Section~\ref{exp}. Both models are trained on the proposed dataset using mask modeling, as detailed in Section~\ref{exp.train}. We use a CFG score of 5 and a rescaling score of 0.75, as stated in Section~\ref{fig:cfg}, while increasing the sampling steps to 100.

The baseline models\footnote{Stable Audio focuses on music generation, so we exclude it from the final T2A comparison but compare its DiT in Section 3.2.} are summarized in Table~\ref{tab:merged_comparison}. To ensure a fair comparison, we use the official checkpoints\footnote{For baselines with multiple versions, we use \textit{tango-full-ft-ac}, \textit{tango-af-ac-ft-ac}, \textit{audioldm-l-full}, and \textit{audioldm2-large}.} or provided samples\footnote{\label{fn:gen-au}Gen-AU releases samples with YouTube IDs but no captions. Since each ID can correspond to five different captions, linking a sample to its original prompt caption isn’t always accurate, which may affect the precision of the CLAP score.} for each baseline. All models are evaluated using the same test method and dataset described in Section~\ref{exp}, following the recommended sampling configurations from their respective papers or repositories.


As shown in Table~\ref{tab:merged_comparison}, 2D U-Net-based models\footnote{\label{fn:audioldm}The open-source AudioLDMs lack fine-tuning or exclusive training on AudioCaps, leading to differences from the paper's best results.} perform worse on FD and IS metrics, producing less realistic audio. Among them, Tango stands out with a strong CLAP score, indicating better coherence.  More recent models\footnote{\label{fn:maa}Make-An-Audio 2 uses all five captions per clip to compute metrics, whereas Tango, AudioLDM, and our method use one randomly selected caption, leading to differences with its reported results.} leveraging a 1D VAE and transformer architecture demonstrate notable improvements in FD and IS. In particular, Gen-AU-Large, benefiting from a larger model scale and extensive pre-training, further enhances audio quality, achieving the highest IS.  EzAudio-L and EzAudio-XL match or surpass baselines across various metrics, highlighting their superior quality and prompt coherence, with EzAudio-XL holding a slight overall advantage.

\begin{figure}[t]
  \centering
  \includegraphics[width=7.5cm]{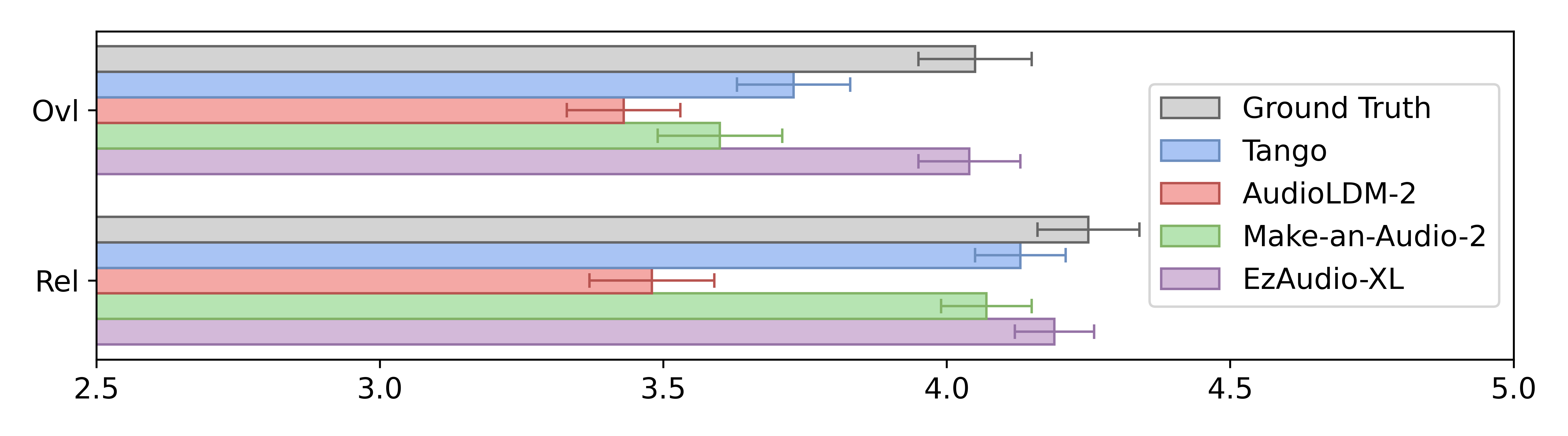}
  \caption{Mean opinion scores with 95\% confidence intervals.}
  \label{fig:sub}
\end{figure}

We conduct a subjective experiment\footnote{Due to cost constraints, we compare EzAudio-XL with Tango, AudioLDM-2, Make-An-Audio-2, and real samples from AudioCaps.} to evaluate overall audio quality (OVL) and text relevance (REL) using a 5-point Mean Opinion Score (MOS) on 30 randomly selected prompts. 12 participants with backgrounds in music production or sound engineering take part.  As shown in Figure~\ref{fig:sub}, results align with objective findings: EzAudio-XL outperforms baselines in both relevance and quality. Make-An-Audio 2 receives a lower OVL score than its objective metrics suggest, likely due to artifacts from synthetic data. Notably, EzAudio-XL’s OVL score approaches real recordings, demonstrating its ability to generate highly realistic audio.


\section{Conclusion}
In this paper, we introduce EzAudio, a framework that integrates a training- and computationally-efficient DiT architecture, an effective training pipeline leveraging synthetic caption data, and CFG rescaling to achieve precise and high-quality audio generation. In future work, we plan to incorporate techniques such as ControlNet, and DPO to further improve controllability and generation quality.

\bibliographystyle{IEEEtran}
\bibliography{mybib}

\begin{thebibliography}{10}
\providecommand{\url}[1]{#1}
\csname url@samestyle\endcsname
\providecommand{\newblock}{\relax}
\providecommand{\bibinfo}[2]{#2}
\providecommand{\BIBentrySTDinterwordspacing}{\spaceskip=0pt\relax}
\providecommand{\BIBentryALTinterwordstretchfactor}{4}
\providecommand{\BIBentryALTinterwordspacing}{\spaceskip=\fontdimen2\font plus
\BIBentryALTinterwordstretchfactor\fontdimen3\font minus \fontdimen4\font\relax}
\providecommand{\BIBforeignlanguage}[2]{{%
\expandafter\ifx\csname l@#1\endcsname\relax
\typeout{** WARNING: IEEEtran.bst: No hyphenation pattern has been}%
\typeout{** loaded for the language `#1'. Using the pattern for}%
\typeout{** the default language instead.}%
\else
\language=\csname l@#1\endcsname
\fi
#2}}
\providecommand{\BIBdecl}{\relax}
\BIBdecl

\bibitem{rombach2022high}
R.~Rombach, A.~Blattmann, D.~Lorenz, P.~Esser, and B.~Ommer, ``High-resolution image synthesis with latent diffusion models,'' in \emph{Proceedings of the IEEE/CVF conference on computer vision and pattern recognition}, 2022, pp. 10\,684--10\,695.

\bibitem{ghosal2023tango}
D.~Ghosal, N.~Majumder, A.~Mehrish, and S.~Poria, ``Text-to-audio generation using instruction tuned llm and latent diffusion model,'' \emph{arXiv preprint arXiv:2304.13731}, 2023.

\bibitem{liu2023audioldm}
H.~Liu, Z.~Chen, Y.~Yuan, X.~Mei, X.~Liu, D.~Mandic, W.~Wang, and M.~D. Plumbley, ``Audioldm: Text-to-audio generation with latent diffusion models,'' in \emph{International Conference on Machine Learning}.\hskip 1em plus 0.5em minus 0.4em\relax PMLR, 2023, pp. 21\,450--21\,474.

\bibitem{huang2023make}
R.~Huang, J.~Huang, D.~Yang, Y.~Ren, L.~Liu, M.~Li, Z.~Ye, J.~Liu, X.~Yin, and Z.~Zhao, ``Make-an-audio: Text-to-audio generation with prompt-enhanced diffusion models,'' in \emph{International Conference on Machine Learning}.\hskip 1em plus 0.5em minus 0.4em\relax PMLR, 2023, pp. 13\,916--13\,932.

\bibitem{evans2024long}
Z.~Evans, J.~D. Parker, C.~Carr, Z.~Zukowski, J.~Taylor, and J.~Pons, ``Long-form music generation with latent diffusion,'' \emph{arXiv preprint arXiv:2404.10301}, 2024.

\bibitem{huang2023make2}
J.~Huang, Y.~Ren, R.~Huang, D.~Yang, Z.~Ye, C.~Zhang, J.~Liu, X.~Yin, Z.~Ma, and Z.~Zhao, ``Make-an-audio 2: Temporal-enhanced text-to-audio generation,'' \emph{arXiv preprint arXiv:2305.18474}, 2023.

\bibitem{vyas2023audiobox}
A.~Vyas, B.~Shi, M.~Le, A.~Tjandra, Y.-C. Wu, B.~Guo, J.~Zhang, X.~Zhang, R.~Adkins, W.~Ngan \emph{et~al.}, ``Audiobox: Unified audio generation with natural language prompts,'' \emph{arXiv preprint arXiv:2312.15821}, 2023.

\bibitem{haji2024taming}
M.~Haji-Ali, W.~Menapace, A.~Siarohin, G.~Balakrishnan, S.~Tulyakov, and V.~Ordonez, ``Taming data and transformers for audio generation,'' \emph{arXiv preprint arXiv:2406.19388}, 2024.

\bibitem{peebles2023scalable}
W.~Peebles and S.~Xie, ``Scalable diffusion models with transformers,'' in \emph{Proceedings of the IEEE/CVF International Conference on Computer Vision}, 2023, pp. 4195--4205.

\bibitem{ho2021classifier}
J.~Ho and T.~Salimans, ``Classifier-free diffusion guidance,'' in \emph{NeurIPS 2021 Workshop on Deep Generative Models and Downstream Applications}, 2021.

\bibitem{chenpixart}
J.~Chen, J.~Yu, C.~Ge, L.~Yao, E.~Xie, Z.~Wang, J.~T. Kwok, P.~Luo, H.~Lu, and Z.~Li, ``Pixart-{\(\alpha\)}: Fast training of diffusion transformer for photorealistic text-to-image synthesis,'' in \emph{The Twelfth International Conference on Learning Representations}, 2024.

\bibitem{mei2024wavcaps}
X.~Mei, C.~Meng, H.~Liu, Q.~Kong, T.~Ko, C.~Zhao, M.~D. Plumbley, Y.~Zou, and W.~Wang, ``Wavcaps: A chatgpt-assisted weakly-labelled audio captioning dataset for audio-language multimodal research,'' \emph{IEEE/ACM Transactions on Audio, Speech, and Language Processing}, 2024.

\bibitem{lin2024common}
S.~Lin, B.~Liu, J.~Li, and X.~Yang, ``Common diffusion noise schedules and sample steps are flawed,'' in \emph{Proceedings of the IEEE/CVF winter conference on applications of computer vision}, 2024, pp. 5404--5411.

\bibitem{chu2023qwen}
Y.~Chu, J.~Xu, X.~Zhou, Q.~Yang, S.~Zhang, Z.~Yan, C.~Zhou, and J.~Zhou, ``Qwen-audio: Advancing universal audio understanding via unified large-scale audio-language models,'' \emph{arXiv preprint arXiv:2311.07919}, 2023.

\bibitem{ghosh2024gama}
S.~Ghosh, S.~Kumar, A.~Seth, C.~K.~R. Evuru, U.~Tyagi, S.~Sakshi, O.~Nieto, R.~Duraiswami, and D.~Manocha, ``Gama: A large audio-language model with advanced audio understanding and complex reasoning abilities,'' \emph{arXiv preprint arXiv:2406.11768}, 2024.

\bibitem{kong2024improving}
Z.~Kong, S.-g. Lee, D.~Ghosal, N.~Majumder, A.~Mehrish, R.~Valle, S.~Poria, and B.~Catanzaro, ``Improving text-to-audio models with synthetic captions,'' \emph{arXiv preprint arXiv:2406.15487}, 2024.

\bibitem{hershey2021benefit}
S.~Hershey, D.~P. Ellis, E.~Fonseca, A.~Jansen, C.~Liu, R.~C. Moore, and M.~Plakal, ``The benefit of temporally-strong labels in audio event classification,'' in \emph{ICASSP 2021-2021 IEEE International Conference on Acoustics, Speech and Signal Processing (ICASSP)}.\hskip 1em plus 0.5em minus 0.4em\relax IEEE, 2021, pp. 366--370.

\bibitem{chung2024scaling}
H.~W. Chung, L.~Hou, S.~Longpre, B.~Zoph, Y.~Tay, W.~Fedus, Y.~Li, X.~Wang, M.~Dehghani, S.~Brahma \emph{et~al.}, ``Scaling instruction-finetuned language models,'' \emph{Journal of Machine Learning Research}, vol.~25, no.~70, pp. 1--53, 2024.

\bibitem{hu2021lora}
E.~J. Hu, Y.~Shen, P.~Wallis, Z.~Allen-Zhu, Y.~Li, S.~Wang, L.~Wang, and W.~Chen, ``Lora: Low-rank adaptation of large language models,'' \emph{arXiv preprint arXiv:2106.09685}, 2021.

\bibitem{bao2023all}
F.~Bao, S.~Nie, K.~Xue, Y.~Cao, C.~Li, H.~Su, and J.~Zhu, ``All are worth words: A vit backbone for diffusion models,'' in \emph{Proceedings of the IEEE/CVF conference on computer vision and pattern recognition}, 2023, pp. 22\,669--22\,679.

\bibitem{le2024voicebox}
M.~Le, A.~Vyas, B.~Shi, B.~Karrer, L.~Sari, R.~Moritz, M.~Williamson, V.~Manohar, Y.~Adi, J.~Mahadeokar \emph{et~al.}, ``Voicebox: Text-guided multilingual universal speech generation at scale,'' \emph{Advances in neural information processing systems}, vol.~36, 2024.

\bibitem{sun2024auto}
L.~Sun, X.~Xu, M.~Wu, and W.~Xie, ``Auto-{ACD}: A large-scale dataset for audio-language representation learning,'' in \emph{ACM Multimedia 2024}, 2024.

\bibitem{li2022blip}
J.~Li, D.~Li, C.~Xiong, and S.~Hoi, ``Blip: Bootstrapping language-image pre-training for unified vision-language understanding and generation,'' in \emph{International conference on machine learning}.\hskip 1em plus 0.5em minus 0.4em\relax PMLR, 2022, pp. 12\,888--12\,900.

\bibitem{wu2023large}
Y.~Wu, K.~Chen, T.~Zhang, Y.~Hui, T.~Berg-Kirkpatrick, and S.~Dubnov, ``Large-scale contrastive language-audio pretraining with feature fusion and keyword-to-caption augmentation,'' in \emph{ICASSP 2023-2023 IEEE International Conference on Acoustics, Speech and Signal Processing (ICASSP)}.\hskip 1em plus 0.5em minus 0.4em\relax IEEE, 2023, pp. 1--5.

\bibitem{gao2023masked}
S.~Gao, P.~Zhou, M.-M. Cheng, and S.~Yan, ``Masked diffusion transformer is a strong image synthesizer,'' in \emph{Proceedings of the IEEE/CVF International Conference on Computer Vision}, 2023, pp. 23\,164--23\,173.

\bibitem{kim2019audiocaps}
C.~D. Kim, B.~Kim, H.~Lee, and G.~Kim, ``Audiocaps: Generating captions for audios in the wild,'' in \emph{Proceedings of the 2019 Conference of the North American Chapter of the Association for Computational Linguistics: Human Language Technologies, Volume 1 (Long and Short Papers)}, 2019, pp. 119--132.

\bibitem{gemmeke2017audio}
J.~F. Gemmeke, D.~P. Ellis, D.~Freedman, A.~Jansen, W.~Lawrence, R.~C. Moore, M.~Plakal, and M.~Ritter, ``Audio set: An ontology and human-labeled dataset for audio events,'' in \emph{2017 IEEE international conference on acoustics, speech and signal processing (ICASSP)}.\hskip 1em plus 0.5em minus 0.4em\relax IEEE, 2017, pp. 776--780.

\bibitem{podell2023sdxl}
D.~Podell, Z.~English, K.~Lacey, A.~Blattmann, T.~Dockhorn, J.~M{\"u}ller, J.~Penna, and R.~Rombach, ``Sdxl: Improving latent diffusion models for high-resolution image synthesis,'' in \emph{The Twelfth International Conference on Learning Representations}, 2024.

\bibitem{hai2024dpm}
J.~Hai, H.~Wang, D.~Yang, K.~Thakkar, N.~Dehak, and M.~Elhilali, ``Dpm-tse: A diffusion probabilistic model for target sound extraction,'' in \emph{ICASSP 2024-2024 IEEE International Conference on Acoustics, Speech and Signal Processing (ICASSP)}.\hskip 1em plus 0.5em minus 0.4em\relax IEEE, 2024, pp. 1196--1200.

\bibitem{liu2024audioldm}
H.~Liu, Y.~Yuan, X.~Liu, X.~Mei, Q.~Kong, Q.~Tian, Y.~Wang, W.~Wang, Y.~Wang, and M.~D. Plumbley, ``Audioldm 2: Learning holistic audio generation with self-supervised pretraining,'' \emph{IEEE/ACM Transactions on Audio, Speech, and Language Processing}, 2024.

\bibitem{kong2020panns}
Q.~Kong, Y.~Cao, T.~Iqbal, Y.~Wang, W.~Wang, and M.~D. Plumbley, ``Panns: Large-scale pretrained audio neural networks for audio pattern recognition,'' \emph{IEEE/ACM Transactions on Audio, Speech, and Language Processing}, vol.~28, pp. 2880--2894, 2020.

\bibitem{su2024roformer}
J.~Su, M.~Ahmed, Y.~Lu, S.~Pan, W.~Bo, and Y.~Liu, ``Roformer: Enhanced transformer with rotary position embedding,'' \emph{Neurocomputing}, vol. 568, p. 127063, 2024.

\end{thebibliography}

\end{document}